# Genome disorder and breast cancer susceptibility

Conor Smyth[1], Iva Špakulová[1], Owen Cotton-Barratt[1], Sajjad Rafiq[2], William Tapper[3], Rosanna Upstill-Goddard[3], John L. Hopper[4], Enes Makalic[4], Daniel F. Schmidt[4], Miroslav Kapuscinski[4], Jörg Fliege[1], Andrew Collins[3], Jacek Brodzki[1], Diana M. Eccles[2], Ben D. MacArthur[1,3,5,*]

[1]Mathematical Sciences, University of Southampton, Southampton SO17 1BJ, UK

[2]Cancer Sciences Academic Unit and University of Southampton Clinical Trials Unit, Faculty of Medicine, University of Southampton and University Hospital Southampton Foundation Trust, Tremona Road, Southampton SO16 6YA, UK

[3]Human Development and Health, Faculty of Medicine, University of Southampton, Tremona Road, Southampton SO16 6YA, UK

[4]Cente for Molecular, Environmental, Genetic and Analytic Epidemiology, School of Population and Global Health, The University of Melbourne, Carlton, VIC, Australia,

[5]Institute for Life Sciences, University of Southampton, Southampton SO17 1BJ, UK

**Abstract**

Many common diseases have a complex genetic basis in which large numbers of genetic variations combine with environmental and lifestyle factors to determine risk. However, quantifying such polygenic effects and their relationship to disease risk has been challenging. In order to address these difficulties we developed a global measure of the information content of an individual's genome relative to a reference population, which may be used to assess differences in global genome structure between cases and appropriate controls. Informally this measure, which we call relative genome information (RGI), quantifies the relative "disorder" of an individual's genome. In order to test its ability to predict disease risk we used RGI to compare single nucleotide polymorphism genotypes from two independent samples of women with early-onset breast cancer with three independent sets of controls. We found that RGI was significantly elevated in both sets of breast cancer cases in comparison with all three sets of controls, with disease risk rising sharply with RGI (odds ratio greater than 12 for the highest percentile RGI). Furthermore, we found that these differences are not due to associations with common variants at a small number of disease-associated loci, but rather are due to the combined associations of thousands of markers distributed throughout the genome. Our results indicate that the information content of an individual's genome may be used to measure the risk of a complex disease, and suggest that early-onset breast cancer has a strongly polygenic basis.

---

* Correspondence to: B.D.MacArthur@soton.ac.uk

**Author Summary**

Recent years have seen significant advances in our understanding of the genetic basis of breast cancer, and a number of markers, such as the BRCA1 and BRCA2 genes, are well-known to be associated with increased disease risk. However, mutations in these genes do not fully account for the genetic basis of breast cancer, and it has proven difficult to identify other powerful markers with high confidence. Accumulating evidence suggests that breast cancer has a multifactorial basis, in which numerous genetic variations combine with environmental and lifestyle factors to determine risk. Here we introduce a simple measure of the information content of an individual's genome and use this measure to show that breast cancer susceptibility is associated with increased genome-wide "disorder", involving subtle alterations in thousands of locations throughout the genome. Our results indicate that information theoretic methods may be used to measure the risk of a complex disease, and suggest that early-onset breast cancer has a strongly polygenic basis.

**Introduction**

Accumulating evidence suggests that many common diseases have a polygenic basis, in which large numbers of genetic variations combine with environmental and lifestyle factors to determine risk (Khoury M.J. 2013). While genome-wide association studies (GWAS), and more recently exome and whole genome sequencing projects, have found hundreds of genetic variants that are associated with disease, the ability to predict susceptibility from these associations is generally low because the contribution of individual variants to risk is often very modest. In the case of breast cancer, the published GWAS have identified markers (single nucleotide polymorphisms, or SNPs) in more than 70 independent regions (loci), the majority with odd ratios less than 1.1 (Bogdanova et al. 2013). Collectively these loci explain, in the statistical but not causative sense, approximately 15% of the familial relative risk which, when combined with the approximately 21% attributed to moderate- to high-penetrance variants (typically very rare mutations) in a dozen or so susceptibility genes, leaves almost two thirds of the familial basis of the disease unaccounted for (Antoniou and Easton 2006; Bogdanova et al. 2013). It is likely that additional genes that explain a proportion of this missing heritability will be found using both whole exome/genome and candidate gene sequencing of familial and young-onset cases, where the genetic component of risk is likely to be greatest (Akbari et al. 2014; Hopper and Carlin 1992; Manolio et al. 2009; Park et al. 2012; Ruark et al. 2013). However, despite such advances, our current understanding of the genetic basis of breast cancer is far from complete.

While most studies to date have focussed on individual genes or gene mutations and their contribution to disease, there has been limited effort to quantify the cumulative effect of variation across the whole genome on disease risk. This is partly due to the historical lack of sufficient data to appropriately quantify normal genomic variation within control populations, and the absence of the statistical techniques needed to analyse such large-scale variation. However, recent years have seen concerted effort to collect and collate the large numbers of genomes (for example the UK Department of Health's 100K initiative http://www.genomicsengland.co.uk) and there is now a need to develop the accompanying methodological tools to assess genomic variation.

In order to begin to address this issue we describe here a measure of the extent to which a set of case genomes differ from a set of control genomes in their global structure. Our method uses ideas from information theory to provide a measure of the information content of an individual's genome with reference to a control population (detailed in Methods below). The essential procedure involves using the reference population to estimate a probability measure on the space of all genomes, and then using the estimated measure to assess how unusual an individual's genome is with respect to the reference population, as quantified by its self-information (also known in information theory as "surprisal") (Cover and Thomas 1991). Formally, the resulting measure, which we refer to as the relative genome information (RGI), is the amount of information, measured in bits, required to specify the observed genome with respect to the unique encoding that minimises the expected number of bits required to specify the genome of an individual drawn at random from the reference population. Informally, the RGI measures how unusual a genome is with respect to the reference population or, since we construct an information-theoretic measure closely related to the Shannon entropy, how "disordered" it is. Thus, someone with a higher RGI has a more unusual genome, either having less common alleles more often than expected, or having some particularly rare alleles. By contrast a lower RGI corresponds to having more common alleles more often, and therefore a less surprising genome.

We hypothesised that global measures of genome variation, such as RGI, might quantify the polygenic basis of complex diseases more completely than GWAS analyses that seek to find statistically significant associations of particular markers with disease. In order to test this hypothesis we compared the RGI of two independent samples of women with early-onset breast cancer genotyped for SNPs relative to three independent samples of unaffected controls (Dite et al. 2003; Eccles et al. 2007; Goodwin et al. 2012; McCredie et al. 1998; Phillips et al. 2005).

## Methods

### Data Normalization

SNP genotypes obtained from blood samples from the following three independent studies were considered: (*i*) The Prospective study of Outcomes in Sporadic versus Hereditary breast cancer (POSH) cohort (Eccles et al. 2007). The POSH cohort consists of approximately 3,000 women aged 40 years or younger at breast cancer diagnosis from which 574 cases were genotyped on the Illumina 660-Quad SNP array. Genotyping was conducted in two batches at separate locations: the Mayo Clinic, Rochester, Minnesota, USA (274 samples) and the Genome

Institute of Singapore, National University of Singapore (300 samples). Only 536 samples that passed quality control filters were considered in this study (Rafiq et al. 2013). (*ii*) The Wellcome Trust Case Control Consortium (WTCCC, http://www.wtccc.org.uk/). The WTCCC consists of approximately 5,000 disease-free controls including individuals from the 1958 British Birth Cohort and from the UK National Blood Service (NBS) Collection. Genotyping was conducted using the Illumina 1.2M chip. (*iii*) The Australian Breast Cancer Family Study (ABCFS) (Dite et al. 2003; McCredie et al. 1998). Cases were a subset of 204 of women aged 40 years or younger at breast cancer diagnosis from the ABCFS; controls were 287 unaffected women aged 40 years and older from the Australian Mammographic Density Twins and Sisters Study (Odefrey et al. 2010). Genotyping was conducted at the Australian Genome Research Facility using the Illumina 610-Quad SNP array.

Only autosomes were considered and SNPs were excluded from each dataset if they failed any of the following quality control filters: minor allele frequencies < 1%; genotyping call rate < 99%; significant deviation from Hardy-Weinberg equilibrium ($p < 0.0001$). All quality control filters were implemented using the software package PLINK (Purcell et al. 2007). In total approximately 475,000 SNPs were genotyped in all five datasets. When comparing datasets and computing RGI only these shared SNPs were considered.

Individuals with evidence of ethnic admixture were excluded by performing multi-dimensional scaling (MDS) analysis using linkage disequilibrium (LD) based pruning ($r^2 > 0.5$) of genotypes in PLINK. This procedure generates a reduced set of approximately independent SNPs. In total there were approximately 133,000 LD-pruned SNPs common to all samples. HapMap data for the African, Asian and Caucasian populations (Gibbs et al. 2003) were also used to provide reference population genotypes against which the genotype data of our cases and controls were compared (Fig. 1a). We identified eight POSH and ten ABCFS samples that showed evidence of mixed ethnicity, and these were excluded from further analysis. Conclusions did not differ without removal of these samples.

**Quantifying Relative Genome Information**

Let $L$ denote a set of locations in the genome (loci), and let $\Lambda = \{A, C, G, T\}$ be the alphabet of possible alleles at each locus $l \in L$. Let $\Pi_l(\lambda, \mu)$ denote the likelihood of finding the unordered allele pair $(\lambda, \mu) \in \Lambda \times \Lambda$ at locus $l \in L$ in the reference population and let $\Pi$ be the product measure of $\Pi_l$ over all $l \in L$. Thus, $\Lambda^{2L}$ denotes the space of all possible genomes, and $\Pi$ represents the probability measure on $\Lambda^{2L}$. Now let $X \in \Lambda^{2L}$ be a genome with allele pair $X_l \in \Lambda \times \Lambda$ at locus $l \in L$. We define the relative local information (RLI) $I_l(X_l) = -\log_2 \Pi_l(X_l)$ at each locus $l \in L$ in the genome $X$ and the relative genome information (RGI) $I(X) = \sum_{l \in L} I_l(X_l)$ for each genome $X$ of interest. For the purposes of comparison it is also convenient to normalize the RGI by $n$, the number of loci genotyped, to give the expected information per locus (EIL), $\mathbb{E}_\Pi(I_l) = \frac{1}{n}\sum_{l \in L} I_l(X_l)$. When comparing sequences of the same length the EIL and RGI are equivalent up to a normalizing factor. However, by normalizing by the number of loci sampled, the EIL allows comparison of relative information content of sequences of different lengths (for instance, comparison of relative information content of different chromosomes). The RLI is the natural information-theoretic measure of the "surprisal" of observing allele pair $X_l \in \Lambda \times \Lambda$ at locus $l \in L$ given the probability measure $\Pi_l$ (Cover and Thomas 1991). Similarly, the RGI is the natural information-theoretic measure of the surprisal of observing the genome $X$, given the probability measure $\Pi$.

In practice $\Pi$ is not known *a priori* and must be estimated from an appropriate reference sample of similar ethnic background to that of the cases. Here, we estimated $\Pi$ using the WTCCC 1958 birth cohort since it was the largest reference sample. In all calculations, $\Pi_l$ was estimated for each locus $l \in L$ using all available genotypes in the reference population at that locus. Once $\Pi$ had been approximated, the RGI was calculated for each genome in each of the remaining four (test) samples (POSH cases, ABCFS cases, ABCFS controls, NBS controls). The two additional independent sets of controls (ABCFS and NBS) were included in order to assess the robustness of the approximation of the background probability measure $\Pi$ from the 1958 cohort alone. For each of the four test samples, missing genotype data at each locus $l \in L$ were assigned the expected value of $\Pi_l$ (i.e. the Shannon entropy $-\sum_{X_l} \Pi_l(X_l) \log_2 \Pi_l(X_l)$ of $\Pi_l$). This method of imputation minimizes the influence of missing data on the calculation of RGI. We also conducted all calculations using only those loci for which there were no missing readings in any of the datasets, and results obtained with and without imputation did not differ qualitatively. A brief worked example illustrating how $\Pi$ was estimated, and the RLI and RGI were calculated, is given in the Supplementary Materials. Estimation of RGI for $N$ case genomes takes $O(n(m + N))$ computational time, where $n$ is the number of loci and $m$ is the number of genomes in the control population, and can be conducted on a desktop PC for moderate sample sizes (thousands of samples and hundreds of thousands of genotyped loci).

**Statistical analysis**

All analysis was conducted in R and Matlab using custom written scripts. The association between EIL and disease odds was estimated using a logistic generalized additive model (Hastie et al. 2009). Tests for significant differences between groups were assessed using Wilcoxon rank-sum tests. All *p*-values were false-discovery rate (FDR) adjusted using the Benjamini-Hochberg procedure (Benjamini and Hochberg 1995).

## Results

We did not observe any difference in expected information per locus (RGI normalized by the number of loci genotyped, EIL) between the three different control sets (1958 controls, NBS controls, ABCFS controls) indicating that the background measure $\Pi$ was reliably estimated; similarly, no difference in EIL between the POSH and ABCFS cases was observed (Fig. 1b-c). However, EIL was significantly higher in both the POSH and ABCFS cases than the three sets of reference controls (FDR adjusted $p < 0.01$, two-sided Wilcoxon rank-sum test) (Fig. 1b-c). Since significant differences within case and control sets were not observed, we amalgamated samples to form one case set (consisting of the ABCFS and POSH cases) and one control set (consisting of the ABCFS, NBS and 1958 controls) for further analysis. Comparison of the distribution of RGI in amalgamated case set and amalgamated control set revealed significant differences in distribution structure ($p = 4.3 \times 10^{-10}$, two-sample Kolmogorov-Smirnov test) with the case distribution having a substantially heavier tail than the control distribution, indicating a greater proportion of samples with higher EIL (Fig. 1d). To investigate further we conducted regression using a logistic generalized additive model (Hastie et al. 2009) in order to estimate the relationship between disease odds ratio and EIL (Fig. 1e). Consistent with the heavy-tailed nature of the case distribution we observed a strong positive association between odds ratio and EIL. In particular, the odds ratio increased sharply for EIL above 1.75, with the highest percentile EIL (above 1.183) having an odds ratio greater than 12 ($p < 1 \times 10^{-16}$, Fisher's exact test). These results indicate that EIL is significantly elevated in breast cancer cases, with the highest percentiles EIL conferring a substantially increased risk.

In order to investigate the genetic basis for these observations we sought to assess whether the differences observed were associated with particular genomic loci or SNP annotations. We began by estimating the number of loci required to account for observed differences at each percentile using random resampling with replacement ($1 \times 10^4$ times) from the case genomes until the required difference was achieved. Differences in median EIL between cases and controls were found to be due to contributions from an estimated 327 distinct loci (95% confidence intervals [306, 349]) (Fig. 1f). The expected number of loci required to account for differences between cases and controls sharply increased with percentile, with differences in the 99th percentile (which conferred the greatest disease risk) requiring an estimated 4954 loci (95% confidence intervals [4921, 5000]) (Fig. 1f). These results indicate that observed differences in EIL are not due to high-penetrance variations at a small number of disease-associated loci, but rather are due to widespread variation at thousands of genomic loci.

In order to investigate this further we assessed the EIL on individual chromosomes. We found that EIL was consistently elevated in the cases by comparison with the controls on 19 of 22 chromosomes (Fig. 2a), and significantly so on 12 of 22 chromosomes (FDR adjusted $p < 0.05$, one-sided Wilcoxon rank-sum test), indicating that differences in EIL are distributed throughout the genome. We also observed notable variations in

EIL by SNP annotation, with the lowest EIL (and therefore the least variation within the samples) occurring in the 5'/3' untranslated and exonic regions, and the highest EIL (and therefore the greatest variation within the samples) occurring in the intergenic regions. This is consistent with previous assessment of relative mutation rates and suggests that 5'/3' UTRs and exonic regions are subject to stronger negative selection than intergenic regions, in accordance with their phenotypic importance (Khurana et al. 2013; Ward and Kellis 2012a, 2012b). In all annotation categories, we again observed a significant increase in EIL in the cases (FDR adjusted $p < 0.05$, one sided Wilcoxon rank-sum test). These results indicate that observed differences in EIL are not localized to distinct regions of the genome (either chromosomes or SNP annotations) but rather are due to widespread variation distributed throughout the genome.

In order to assess whether increased genomic disorder affects disease prognosis we also looked for associations between EIL and Nottingham Prognostic Index (NPI) (Galea et al. 1992) and age at diagnosis within the POSH cases (corresponding information for the ABCFS cases was not available). However, we found no significant associations (Fig. 2d-e) suggesting that, within the power of this study, EIL is not a predictor of outcome or age of onset although, due to the small number of samples with complete information, this possibility cannot be excluded.

Taken together, our results indicate that genome-wide disorder is positively associated with breast cancer risk. Prior to analysis all genotyping data were subjected to stringent quality assurance and we observed no association between sex, sequencing platform, time/place of sequencing and EIL, indicating that poor data quality or variation in genotype due to ethnicity or sex are unlikely to explain our results (Fig. 1b-c, Fig. 2f). Rather, changes in EIL appear to quantify biologically meaningful differences in large-scale genome structure between breast cancer cases and controls.

**Discussion**

Many of the currently known moderate- and high-risk cancer susceptibility genes code for proteins involved in DNA repair, and DNA damage signatures found in tumour tissue may reflect a core component of the underlying aetiology of cancer (Nik-Zainal et al. 2012a; Nik-Zainal et al. 2012b; Stephens et al. 2012). Thus, DNA repair deficiencies may have a role in elevating the EIL in cancer cases. Although signs of tumour-associated DNA repair deficiencies are not typically found in normal blood derived DNA samples, two recent reports indicate that inherited DNA repair deficiency can be detected in DNA extracted from peripheral blood lymphocytes (Ingham et al. 2013; Ruark et al. 2013). It is possible that rapid cellular replication subsequent to chemotherapy-induced myelosupression could amplify such signals. However, we found no evidence of association between EIL and chemotherapy status (Fig. 2c), suggesting that this is not the case. An alternative possibility is that variation in genome-wide disorder (and therefore EIL) arises due to polymorphisms in genes associated with the DNA damage repair pathways that safeguard germline integrity during oogenesis (Kerr et al. 2012; Levine et al. 2011; Suh et al. 2006). By regulating the fidelity with which genomic information is transmitted between generations such polymorphisms could generate variations in the rate of production of *de novo* mutations in successive generations that affect susceptibility to disease in later life. While we have not tested this hypothesis directly, our results are consistent with this perspective. In future familial studies it would be interesting to assess whether parental mutations in the p53/p63/p73 pathways (which have a central role in protection of the maternal and paternal genomes) (Levine et al. 2011) are associated with increased genomic disorder in their offspring.

Taken together our analysis indicates that early-onset breast cancer has a strongly polygenic basis, involving variation at thousands of markers distributed throughout the genome. Thus, along with assessment of known risk-associated variants, the information content of an individual's genome is a useful predictor of disease susceptibility. Further analysis of the relationship between global genome structure and disease risk may reveal a similarly polygenic basis for a variety of other complex diseases.

**Figure legends**

**Figure 1.** (**a**) Multidimensional scaling plot of all samples and HapMap2 populations genotyped for ~133,000 SNPs. (**b**) Expected information per locus (EIL) for each of the different datasets. Median ± 95% confidence intervals are shown. (**c**) Matrix of FDR adjusted *p*-values for comparisons of medians (two-sided Wilcoxon rank-sum test). (**d**) Q-Q plot of EIL in cases versus controls. *p*-value from a two-sample Kolmogorov-Smirnov test is shown. (**e**) Estimated odds ratio as a function of EIL. (**f**) The number of loci that account for the differences in EIL observed between cases and controls by percentile. 95% confidence intervals are within the markers, so are not shown.

**Figure 2.** (**a**) Expected information per locus (EIL) by chromosome. Median ± 95% confidence intervals are shown. Stars indicate significant changes at FDR adjusted $p < 0.05$ by one-sided Wilcoxon rank-sum test. (**b**) Relationship between EIL and chemotherapy status in the POSH cohort. (**c**) Relationship between EIL and Nottingham Prognostic Index (NPI) in the POSH cohort. (**d**) Relationship between EIL and age at diagnosis in the POSH cohort. In panels (**b-d**) *p*-values for Spearman tests for association are shown. (**f**) EIL in males and females in the controls. Median ± 95% confidence intervals are shown.

**Acknowledgements**

This work was supported by Engineering and Physical Sciences Research Council grant EP/I016945/1 and a Breast Cancer Campaign PhD studentship.

**Author contributions**

BDM, CS, OC-B, JF, JB and DE conceived and designed the analysis. SR, WT, RU-G, EM, DFS, MK and AC processed the genotype data, and performed all normalization. BDM, IS, CS, and OC-B performed the RGI analysis. DE and JH provided genotype data. BDM, JH and DE interpreted the results. BDM, CS and O-CB wrote the paper, with input from all other authors. All authors approved the submission of the manuscript.

**References**

Akbari, M. R., et al. (2014), 'PPM1D Mutations in Circulating White Blood Cells and the Risk for Ovarian Cancer', *Jnci-Journal of the National Cancer Institute,* 106 (1).

Antoniou, A. C. and Easton, D. F. (2006), 'Models of genetic susceptibility to breast cancer', *Oncogene,* 25 (43), 5898-905.

Benjamini, Y. and Hochberg, Y. (1995), 'Controlling the False Discovery Rate - a Practical and Powerful Approach to Multiple Testing', *Journal of the Royal Statistical Society Series B-Methodological,* 57 (1), 289-300.

Bogdanova, N., Helbig, S., and Dork, T. (2013), 'Hereditary breast cancer: ever more pieces to the polygenic puzzle', *Hereditary Cancer in Clinical Practice,* 11.

Cover, T.M. and Thomas, J.A. (1991), *Elements of Information Theory* (New York: Wiley and Sons).

Dite, G. S., et al. (2003), 'Familial risks, early-onset breast cancer, and BRCA1 and BRCA2 germline mutations', *Journal of the National Cancer Institute,* 95 (6), 448-57.

Eccles, D., et al. (2007), 'Prospective study of Outcomes in Sporadic versus Hereditary breast cancer (POSH): study protocol', *BMC Cancer,* 7, 160.

Galea, M. H., et al. (1992), 'The Nottingham Prognostic Index in Primary Breast-Cancer', *Breast Cancer Research and Treatment,* 22 (3), 207-19.

Gibbs, R. A., et al. (2003), 'The International HapMap Project', *Nature,* 426 (6968), 789-96.

Goodwin, P. J., et al. (2012), 'Breast Cancer Prognosis in BRCA1 and BRCA2 Mutation Carriers: An International Prospective Breast Cancer Family Registry Population-Based Cohort Study', *Journal of Clinical Oncology,* 30 (1), 19-26.

Hastie, T, Tibshirani, R, and Friedman, J (2009), *The elements of statistical learning* (New York: Springer).

Hopper, J. L. and Carlin, J. B. (1992), 'Familial Aggregation of a Disease Consequent Upon Correlation between Relatives in a Risk Factor Measured on a Continuous Scale', *American Journal of Epidemiology,* 136 (9), 1138-47.

Ingham, D., et al. (2013), 'Simple detection of germline microsatellite instability for diagnosis of constitutional mismatch repair cancer syndrome', *Human Mutatation,* 34 (6), 847-52.

Kerr, J. B., et al. (2012), 'DNA Damage-Induced Primordial Follicle Oocyte Apoptosis and Loss of Fertility Require TAp63-Mediated Induction of Puma and Noxa', *Molecular Cell,* 48 (3), 343-52.

Khoury M.J., Janssens A.C.J.W, Ransohoff David F. (2013), 'How can polygenic inheritance be used in population screening for common diseases?', *Genetics in Medicine,* 15, 437-43.

Khurana, E., et al. (2013), 'Integrative Annotation of Variants from 1092 Humans: Application to Cancer Genomics', *Science,* 342 (6154), 1235587.

Levine, A. J., et al. (2011), 'The p53 family: guardians of maternal reproduction', *Nature Reviews Molecular Cell Biology,* 12 (4), 259-265.

Manolio, T. A., et al. (2009), 'Finding the missing heritability of complex diseases', *Nature,* 461 (7265), 747-53.

McCredie, M. R. E., et al. (1998), 'Breast cancer in Australian women under the age of 40', *Cancer Causes & Control,* 9 (2), 189-98.

Nik-Zainal, S., et al. (2012a), 'The Life History of 21 Breast Cancers', *Cell,* 149 (5).

Nik-Zainal, S., et al. (2012b), 'Mutational Processes Molding the Genomes of 21 Breast Cancers', *Cell,* 149 (5), 979-93.

Odefrey, F., et al. (2010), 'Common Genetic Variants Associated with Breast Cancer and Mammographic Density Measures That Predict Disease', *Cancer Research,* 70 (4), 1449-58.

Park, D. J., et al. (2012), 'Rare Mutations in XRCC2 Increase the Risk of Breast Cancer', *American Journal of Human Genetics,* 90 (4), 734-39.

Phillips, K. A., et al. (2005), 'Agreement between self-reported breast cancer treatment and medical records in a population-based breast cancer family registry', *Journal of Clinical Oncology,* 23 (21), 4679-86.

Purcell, S., et al. (2007), 'PLINK: A tool set for whole-genome association and population-based linkage analyses', *American Journal of Human Genetics,* 81 (3), 559-75.

Rafiq, S., et al. (2013), 'Identification of inherited genetic variations influencing prognosis in early-onset breast cancer', *Cancer Res,* 73 (6), 1883-91.

Ruark, E., et al. (2013), 'Mosaic PPM1D mutations are associated with predisposition to breast and ovarian cancer', *Nature,* 493 (7432), 406-10.

Stephens, P. J., et al. (2012), 'The landscape of cancer genes and mutational processes in breast cancer', *Nature,* 486 (7403), 400-04.

Suh, E. K., et al. (2006), 'p63 protects the female germ line during meiotic arrest', *Nature,* 444 (7119), 624-28.

Ward, L. D. and Kellis, M. (2012a), 'Evidence of Abundant Purifying Selection in Humans for Recently Acquired Regulatory Functions', *Science,* 337 (6102), 1675-78.

--- (2012b), 'Interpreting noncoding genetic variation in complex traits and human disease', *Nature Biotechnology,* 30 (11), 1095-106.

**Supplementary Materials**

**A worked example**

Consider sampling 5 individuals from an unknown background population. In this illustrative example there are 2 letters in the alphabet $\Lambda = \{A, B\}$ and at a particular locus the following alleles are observed:

|         | Individual 1 | Individual 2 | Individual 3 | Individual 4 | Individual 5 |
|---------|--------------|--------------|--------------|--------------|--------------|
| Locus 1 | AA           | AA           | AA           | BB           | BA           |

To estimate the true probability of finding each allele pair in the background population we count the frequencies of the alleles A and B to obtain:

$$p(A) = \frac{7}{10}, \quad p(B) = \frac{3}{10}.$$

We may then estimate the Hardy-Weinberg proportions:

$$\Pi_1(AA) \approx p(A) \times p(A) = \left(\frac{7}{10}\right)^2,$$

$$\Pi_1(BB) \approx p(B) \times p(B) = \left(\frac{3}{10}\right)^2,$$

$$\Pi_1(AB) = \Pi_1(BA) \approx 2 \times p(A) \times p(B) = 2 \times \left(\frac{7}{10}\right) \times \left(\frac{3}{10}\right).$$

The RLI for each pair is found by taking the negative logarithm of these probabilities:

$$-\log_2(\Pi_1(AA)) = 1.0291,$$

$$-\log_2(\Pi_1(BB)) = 3.4739,$$

$$-\log_2(\Pi_1(AB)) = 1.2515,$$

where logarithms are taken to base 2 so that information is measured in bits. Since the pair BB is rare in the population it imparts more information whenever it is observed; by contrast since the pair AA is common, it imparts less information whenever it is observed.

Suppose now that a second locus is also sequenced and the following alleles are observed:

|        | Individual 1 | Individual 2 | Individual 3 | Individual 4 | Individual 5 |
|--------|--------------|--------------|--------------|--------------|--------------|
| Locus 2 | AA | BA | BA | BB | AA |

A similar calculation yields the RLI of the second locus and the RGI for each individual (here, the 'genome' has just 2 loci):

|       | Individual 1 | Individual 2 | Individual 3 | Individual 4 | Individual 5 |
|-------|--------------|--------------|--------------|--------------|--------------|
| RLI 1 | 1.0291 | 1.0291 | 1.0291 | 3.4739 | 1.2515 |
| RLI 2 | 1.4739 | 1.0589 | 1.0589 | 2.6439 | 1.4739 |
| RGI   | 2.5030 | 2.0880 | 2.0880 | 6.1178 | 2.7254 |

In this case, individual 4 has the least common alleles at both loci, and correspondingly has the largest RGI. Individuals 2 and 3 have the most common alleles are therefore the lowest RGI. In this calculation the five individuals form the control population from which the probability measure $\Pi$ on $\Lambda^{2L}$ is approximated. Once this measure has been approximated, it may be used to assess the RGI of unseen case genomes. For example, a case individual with the genome [AA, BB] has RGI = 1.0291 + 2.6439 = 3.673.

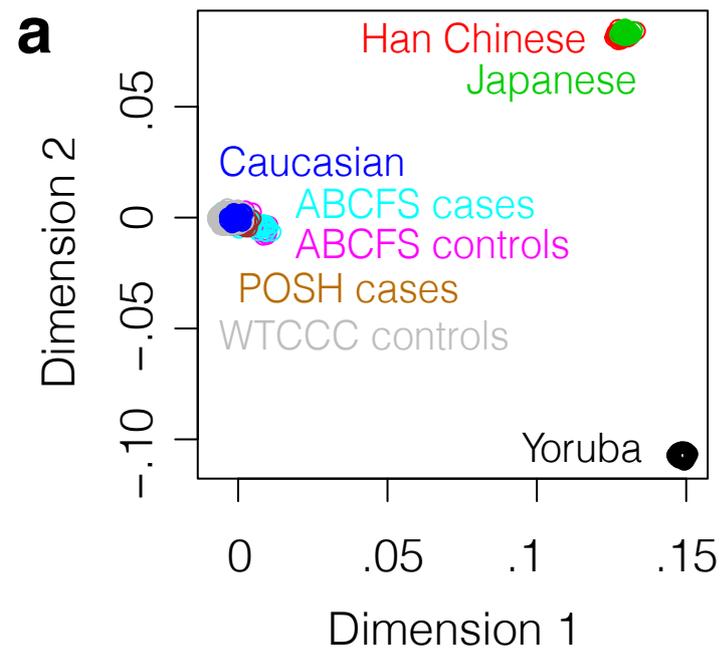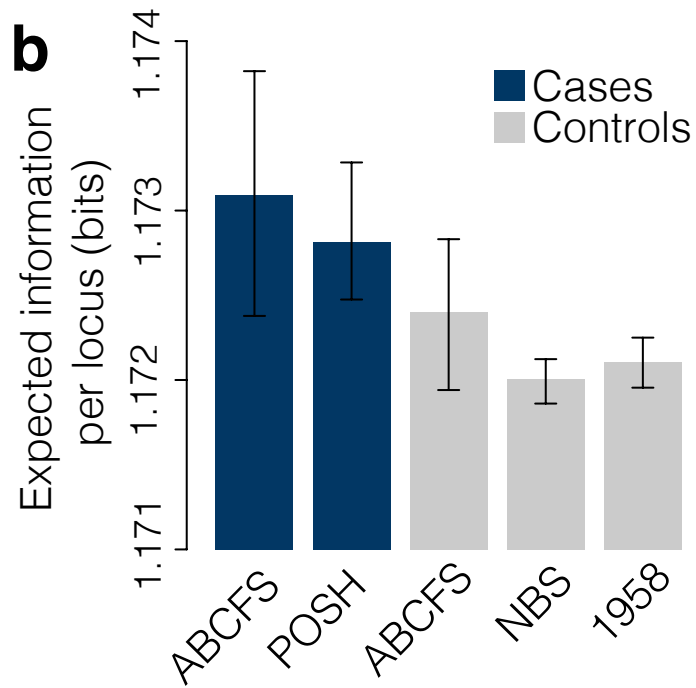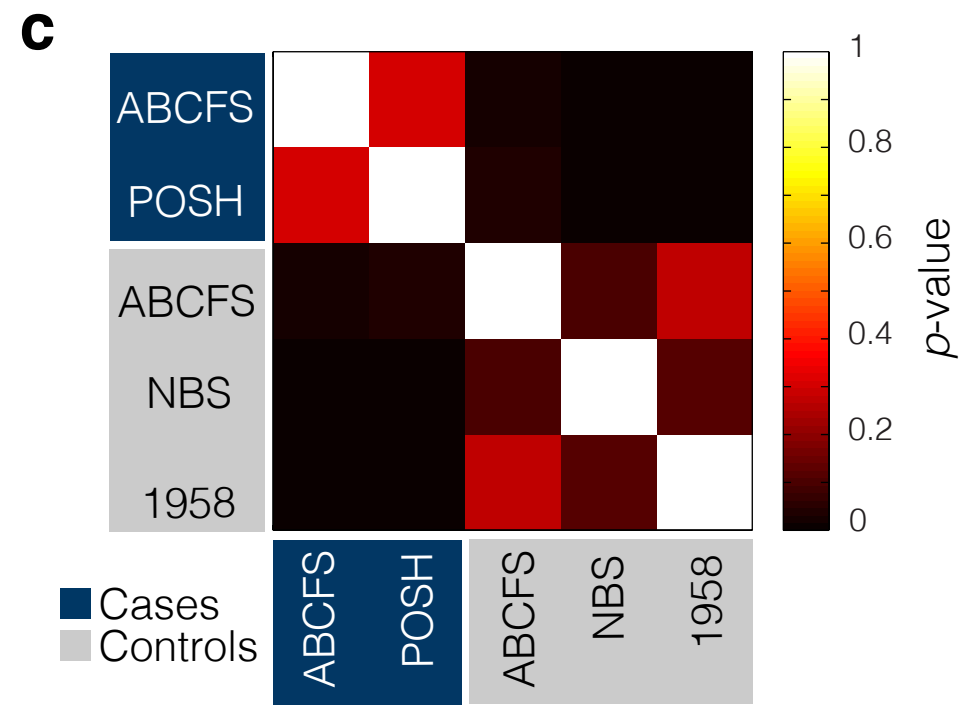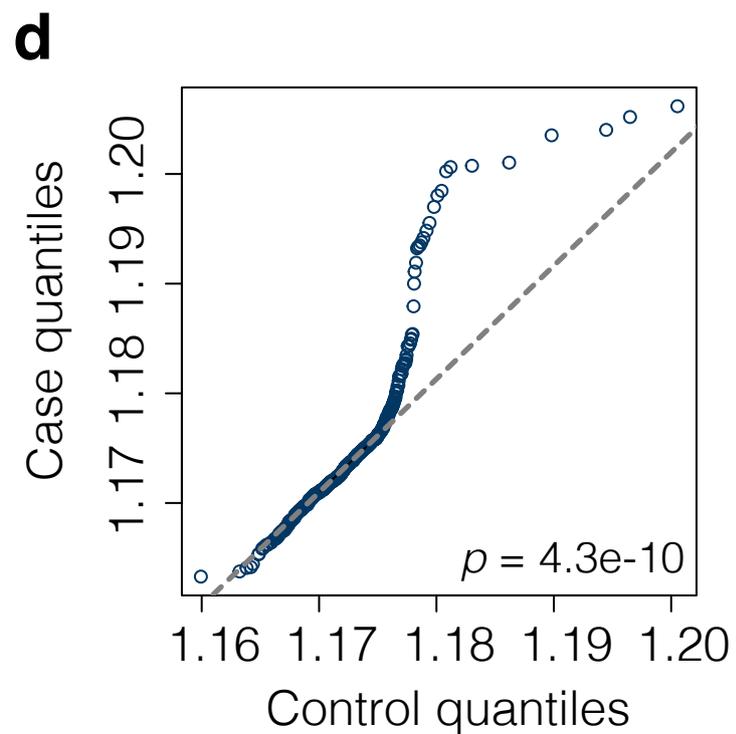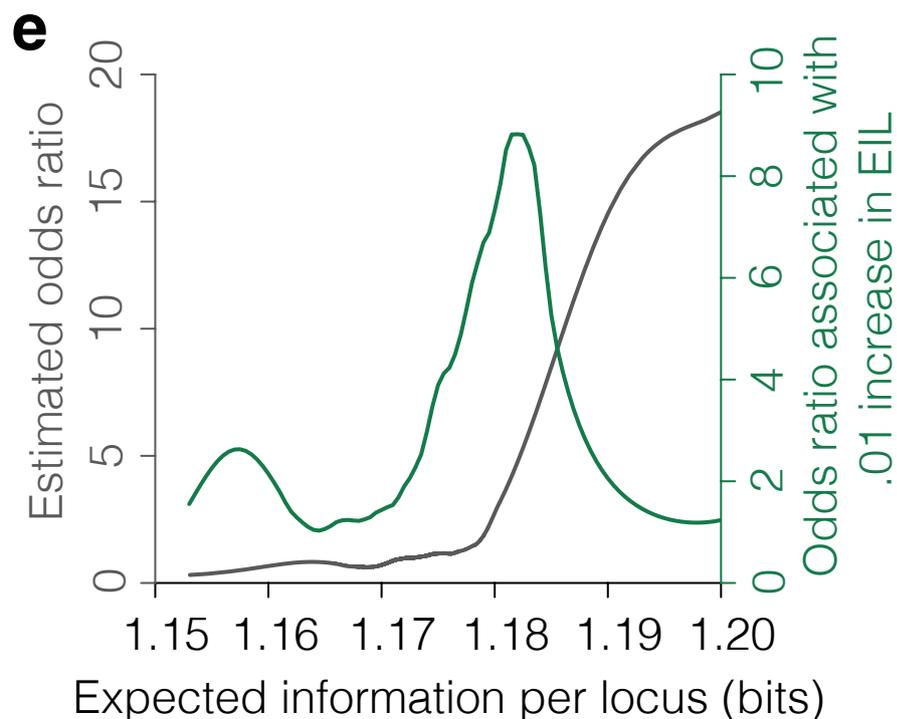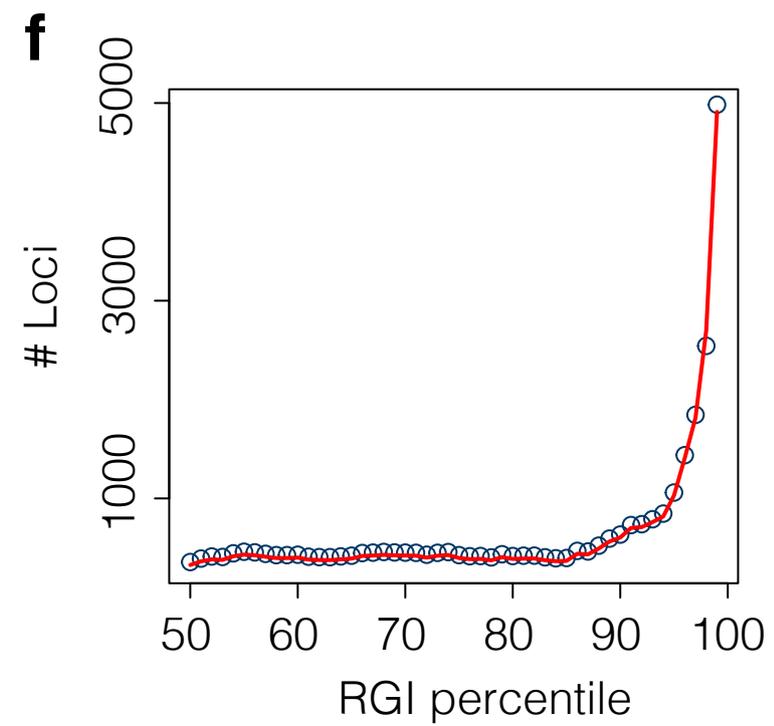

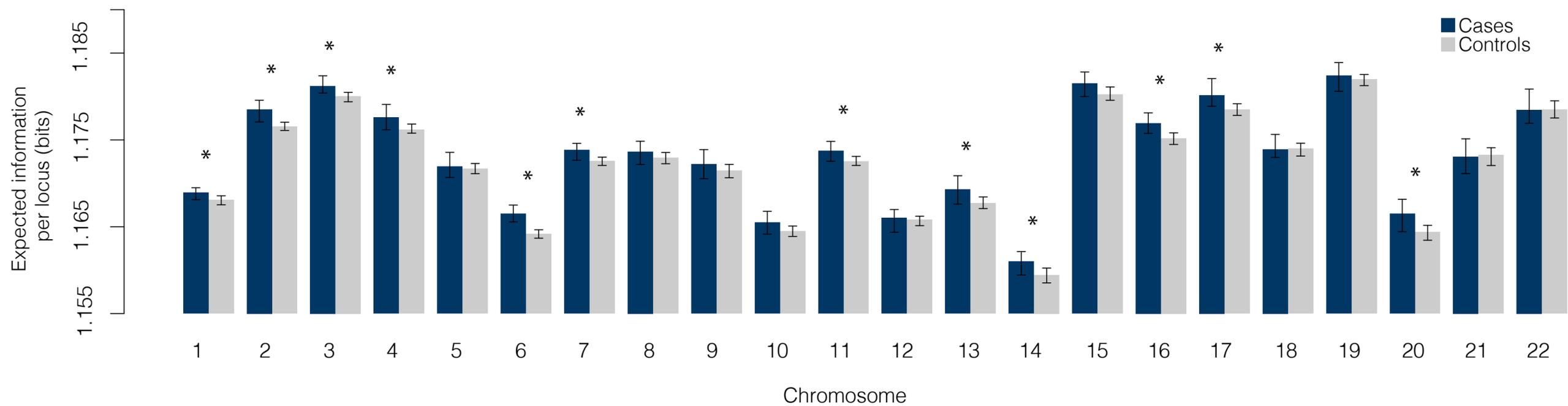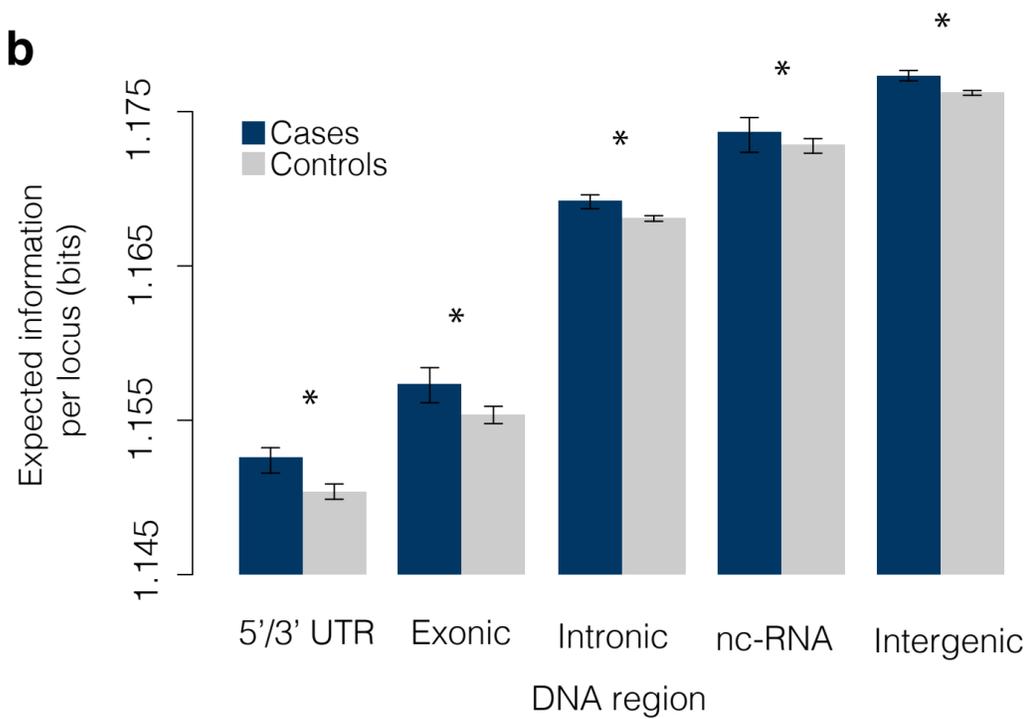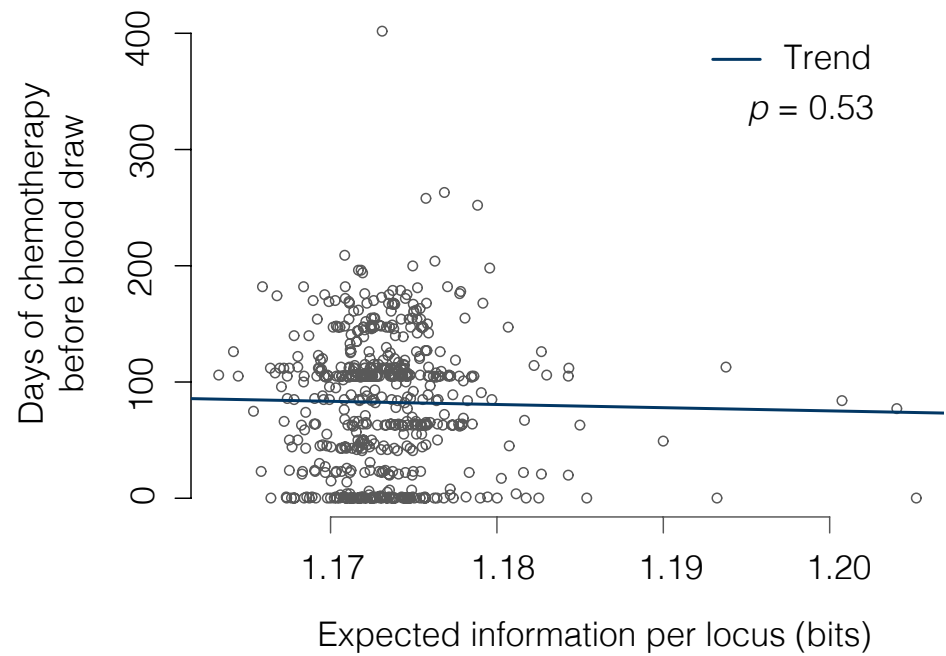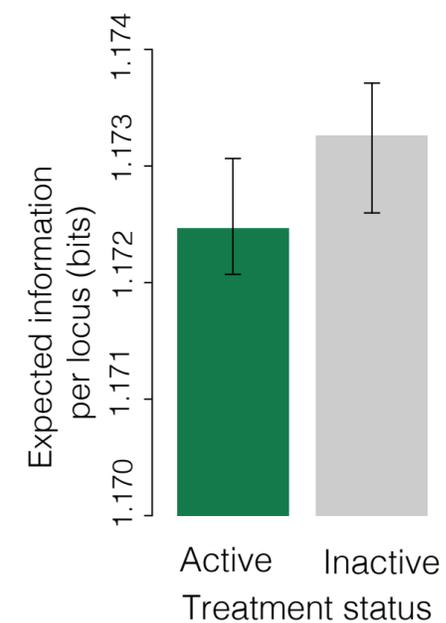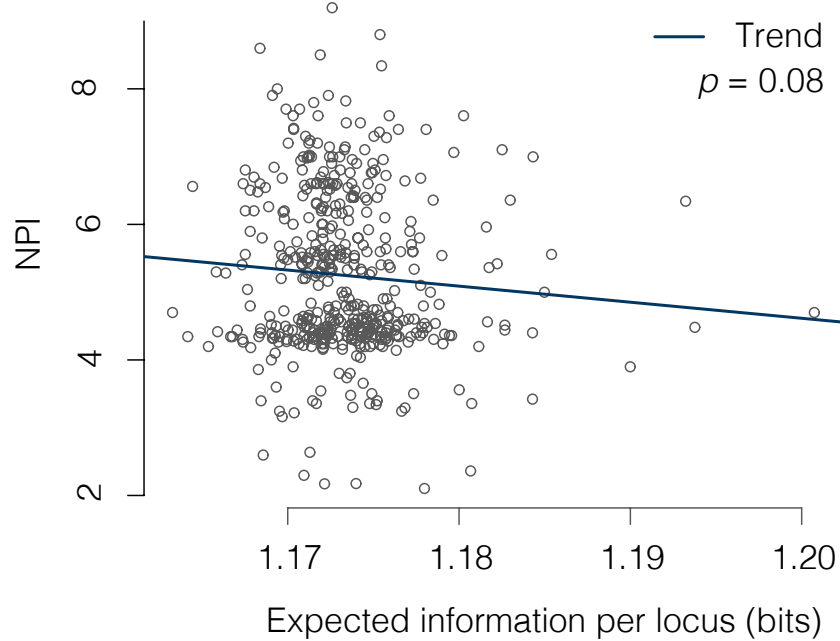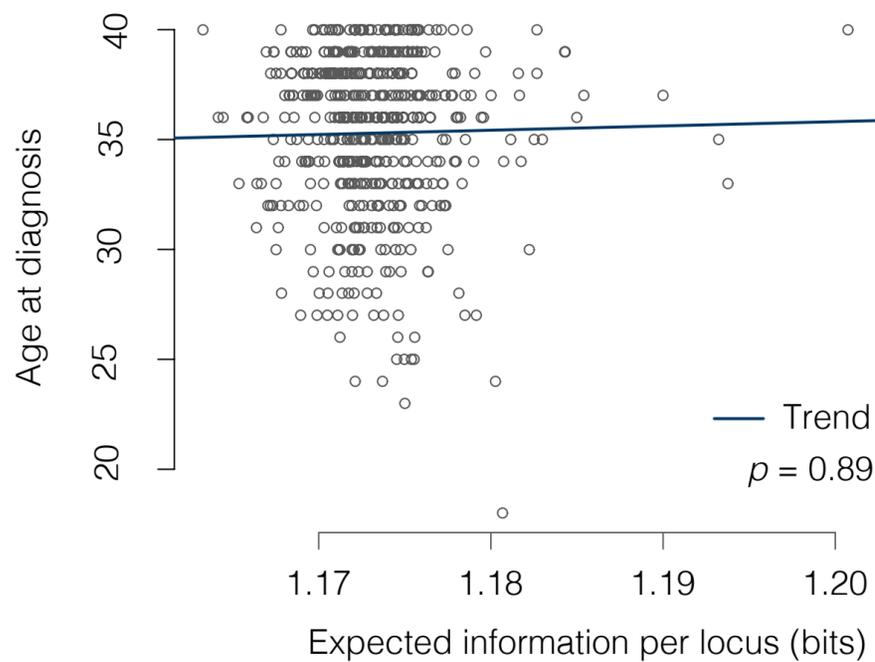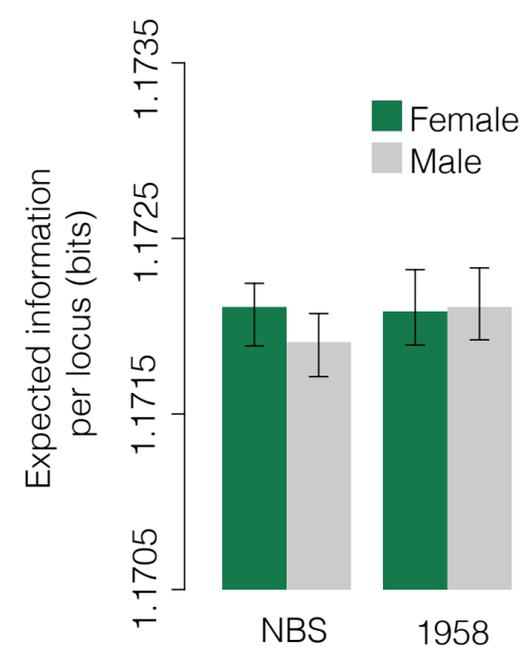